\documentstyle[12pt]{article}
\title  {Two-Body Mass-Shell Constraints \\
in a Constant Magnetic Field\\
 (Neutral  Case)}

\medskip

\author { Philippe Droz-Vincent\\[2mm]
Laboratoire de Gravitation et Cosmologie Relativistes\\
C.N.R.S./ ESA 7065, Universit\'e Pierre et Marie Curie\\
Tour 22-12 , boite courrier 142,\\
 4 place Jussieu 75252 Paris Cedex 05, France}
\date {02.07.99}
\newcommand{\beq}{\begin{equation}}
\newcommand{\eeq}{\end{equation}}
\newcommand{\alp}{\alpha}    
 \newcommand {\half}{ {1 \over 2}} 
 \newcommand {\noi}{\noindent}     
\newcommand {\st}{_{\scriptscriptstyle T} }
\newcommand {\longi}{_ {\scriptscriptstyle L}  } 
\newcommand {\zero}{_ {(0)} } 
\newcommand {\zer}{^  {(0)} } 
\newcommand  {\per}{ _ \perp }

\newcommand {\lam}{\lambda}
\newcommand  {\disp}{\displaystyle}

\begin{document}
\maketitle
\abstract{A constant homogeneous magnetic field is applied to a composite 
system made of two scalar particles with opposite charges.
Motion is described by a pair of coupled Klein-Gordon equations that are 
written in closed form with help of a suitable representation.
The relativistic symmetry associated with the magnetic field is carefully 
respected. 
Considering eigenstates of the pseudomomentum four-vector,
we separate out collective  variables and obtain a three-dimensional reduced 
equation, posing a nonconventional eigenvalue problem.
 The velocity of the system as a whole (with respect to the frames where 
the field is purely magnetic) generates "motional terms" in the formulas;
these terms are taken into account within a  manifestly covariant framework.}

 \bigskip

\vfill \eject

\section {Introduction}
 
    The theory of many-particle systems in external fields requires 
particular caution, even in the simple framework of nonrelativistic mechanics:
as soon as all the constituent masses are of comparable magnitudes, it 
becomes difficult to disentangle the  dynamics of relative variables from 
the motion of the center of mass.

\noi       The case of a globally neutral system of charges imbedded in a 
constant homogeneous magnetic field is of special interest however, because 
(under very general assumptions) it enjoys this property that  the total 
{\it pseudomomentum} ${\vec C} =  \sum {\vec p} + e {\vec A}$
 is conserved and has mutually commuting components
\cite{3}\cite{dzy}\cite{gro} \cite{avr}. 
This exceptional circumstance permits to separate, in a generalized sense, 
relative motion, and therefore provides a clean-cut definition of what is the 
spectrum of the system \cite{avr}.

\noi    Relativistic corrections have soon been considered \cite{gro} in a 
three-dimensional framework; this is certainly sufficient in a  large number of 
applications,  but fails to account for  the  relativistic symmetry.
  Indeed the constant magnetic field has this peculiarity that it does not
 correspond to a unique "laboratory  frame". 
When a constant homogeneous  electromagnetic field  is seen
as purely magnetic in some frame (conventionally referred to as lab frame), 
such a frame cannot be unique \cite{DVNCim}, thus    total energy, 
if defined as the (conserved) time component of  linear momentum,
 is affected by this ambiguity.  
All the directions eligible for the time axis of a possible lab frame span a
 two-dimensional  plane $(E\longi)$ with hyperbolic metric;   so we are led to
pay attention to  special Lorentz transformations in this "longitudinal plane".
                              
\noi  Thus a four-dimensional  spacetime  approach is warranted in order
 to keep under control the full relativistic symmetry of motion.

\medskip
\noi In this paper we focus on two-body systems, because the covariant methods 
of relativistic particle dynamics are  well  understood and more tractable in
 this case.
In previous works \cite{DVNCim}\cite{dv95}  we have indicated  how the 
mass-shell 
constraints for  two scalar particles undergoing mutual interaction can be
 minimally coupled (in closed form and remaining compatible)  with an external 
electromagnetic field  $F_{\mu \nu}$ wich can be  either pure electric or pure 
magnetic.                          
In both cases a four-vector  $C^\alp$, called pseudomomentum,
 is conserved and for {\it neutral 
systems} its four components commute among themselves.   
Writting down explicit equations of motion  requires that we go  to a new
 representation, adapted to the symmetries of the external field.

\noi  When, as we assume here,    $F_{\mu \nu}$  is purely
 magnetic \cite{purmagn},
  a further change of representation eliminates not only  the
 collective variables conjugate to pseudomomentum  but also a fifth 
 variable  which is nothing but relative time.
The outcome is a manifestly covariant  equation  to be solved for a  reduced 
wave function which depends only  on three spacelike degrees of freedom.
 The material that we published so far \cite{DVNCim} 
was limited  to the general lines of this approach.

\noi  In this article  we  explicitly carry out the change of representation 
and write down  the reduced wave equation in a tractable  form,
 showing   the  details  of the   various  contributions it contains.
In addition we  discuss whether the reduced wave equation can be considered as 
an eigenvalue problem, and for which parameter.
We prepare an eventual perturbation theory which will ultimately  result in
 a covariant framework for the spectroscopy of two-body systems. 

\noi  In Section 2 we display   the notation used and we remind several results 
from previous works.
Section 3 is devoted to the explicit reduction of the number of degrees of 
freedom, and to a qualitative discussion about the various terms arising in 
the reduced wave equation. Section 4 contains concluding remarks and an 
outlook.

\medskip
\noi 

\bigskip
\noi \section{Basic Equations, Symmetries}

\noi
When pair creation can be neglected, a system of two scalar particles can be
 described by a pair of coupled Klein-Gordon equations                    
$$ \displaystyle 
2 H_a   \Psi = m _a ^ 2  \  \Psi  \qquad \quad  a,b = 1,2     $$
referred to as the {\it mass-shell constraints\/}.
 Here $\Psi $ has two arguments $q_1, q_2$ running in spacetime.
We  cover all cases of practical interest assuming that 
$ H _a = K_a + V $.

\noi        In the above  formula  $2K_a =  (p_a - e_a  A (a) ) ^2$
 is the squared-mass operator for particle $a$ alone in the magnetic field,
 and $V$ is a suitable modification of the term  $ V \zer $
 which would describe the mutual interaction in the absence of external field;
 this modification is necessary in order to 
keep the mass-shell constraints mutually compatible when the  field 
$F_{\alp \beta} $  is applied.

\noi         For all vectors
        $\xi , \eta $ we write 
   $\xi \cdot    F  \cdot \eta $ for   
   $\xi ^\alpha     F _{\alpha \beta}   \eta ^\beta  $. With a similar 
convention  
 $  \disp   A(a) = \half  \     q_a  \cdot  F $
in a  Lorentz-covariant gauge.  

\noi   Notice that $\xi \cdot \eta \longi =  \xi \longi \cdot \eta \longi $ 
and    $\xi \cdot \eta \st  =  \xi \st  \cdot \eta \st  $.

    \noi
An important technical point is that  applying a constant magnetic field
 provides a unique and invariant decomposition of any four-vector $\xi$ into
 longitudinal and transverse parts, say 
  $ \xi = \xi \longi  + \xi \st $.   The orthocomplement of $(E\longi)$ in the 
space of four-vectors  is a  two-dimensional  plane $(E\st)$ endowed with 
elliptic metric. In any adapted frame, $\xi \longi$ (resp. $\xi \st$) 
has nonvanishing coordinates $\xi ^0 , \xi ^3 $ (resp. $ \xi^1 , \xi^2$).

\medskip

\noi  The theory of relativistic two-body systems,  formulated  many  years
 ago along the lines of "predictive mechnics" and "constraints theory" 
\cite{rep} \cite{tod} \cite{barcelo} \cite{coupl}  
has been more recently extended to  cases where some  external field is present
\cite {bij89} \cite{DVNCim}. Here we 
assume that a constant homogeneous magnetic field
 is applied to a pair of opposite charges, say       $e_1 =-e_2 = e$.

\medskip
\noi        The constraint approach employed here has over the
 Bethe-Salpeter  equation 
several advantages; for example in the particular case of isolated systems
 (no field applied)   the dependence on relative 
time  gets automatically factorized out  \cite{sazdj}.

\noi It is convenient to re-arrange the canonical variables as follows
$$ z= q_1 -q_2    \qquad \qquad               Q= \half (q_1 + q_2)    $$
$$ y = \half  \   (p_1 -p_2) \qquad  \qquad     P = p_1 + p_2         $$
so $ [z,y]= [Q,P]=-i \delta $, etc.

\noi  The Lie algebra of the Lorentz group is generated by the tensor 
$$      M   =    M_1  + M_2     =     Q\wedge P + z\wedge y      $$
with  $M_1    = q_1\wedge p_1 , \quad   M_2 =   q_2 \wedge p_2$. 
In any adapted frame, rotations in $(E \st)$ are   generated by $M_{12}$ 
and boosts   in $(E \longi )$ by $M_{03}$. 

\noi  
An essential ingredient of mutual interactions  \cite{rep} is the quantity  
$\disp   {\widetilde z}^2 = z^2 - (z \cdot P )^2 / P^2 $.
But in order to avoid denominators in calculations, 
it is  convenient to  employ 
  \beq  Z = z^ 2  P ^ 2  -  ( z \cdot  P )^ 2    \label{2.1}  \eeq
We shall assume   that 
      \beq  V\zer = f(Z, P^ 2, y \cdot  P)   \label{2.2}   \eeq
This form is general enough to accomodate a large class of interactions.

\medskip
\noi  {\sl Definition}

\noi 
When speaking of {\em energy-dependent interactions}, we refer to the {\em 
total} energy of isolated systems, namely $\sqrt {P^2}$.

\noi               Although $Z$ is more 
practical for calculations, it would be more  natural to take 
 ${\widetilde z}^2$ and $P^2$ as  independent dynamical variables, defining
$\disp   g ( Z/P^2, P^2 , y \cdot P)   =   f(Z, P^ 2, y \cdot  P) $.
Therefore we  say  that $V\zer$ does{\em not} depend on (total) energy when
the function $f$ takes on the form
$f = g(Z/P^2, y \cdot P) $.

\medskip   \noi
Although  $f$ in (\ref{2.2}) is supposed to be known, it would be a problem 
to determine $V$ in closed form.
In the  external-field representation, which involves  a new wave function
 $\Psi '$ and new operators $H_a ',  K_b ', V' $, this problem is solved by 
making the ansatz
\beq    V^\prime  =  f( \widehat Z, P^2, y\longi .P \longi )   
     \label{2.3}  \eeq
where 
$ \widehat Z= Z' \zero=    (Z')_   {F=0}     $ (it turns out that 
$     \widehat Z $ commutes with $  y\longi \cdot P \longi $).
The explicit form of $ \widehat Z $ was calculated in ref. \cite{DVNCim}.
      \beq \widehat Z =
Z  +  2(P\longi ^2 \   z \cdot P - P^2  \   z\longi \cdot P\longi  ) L
  +  P\st^2 P\longi ^2  L^2                    \label{2.4}  \eeq
   where the scalar  $L$ is defined as 
 \beq  L = {P \longi \cdot z   \over  (P \longi )^ 2}   \label{2.5} \eeq
The equations of motion are compatible provided  \cite{DVNCim} that 
 $ \widehat Z $ commutes with $y\longi \cdot P \longi $.
 Let us transform (\ref{2.4}) in order to render this  commutation 
property manifest.
First we split  $z$ as the sum of $z\longi $ and $z \st $ in   $Z$, hence
\beq Z = (z\st ^ 2 + z \longi ^ 2 ) P ^2  
- (z \st \cdot P )^ 2 - ( z\longi \cdot P )^ 2
-2 (z \st \cdot P )(z \longi \cdot P )            \label{2.6}      \eeq
Develop (\ref{2.4}) and perform  elementary manipulations
 using (\ref{2.6}). We get
$$ \widehat Z = Z + 2 (z \st \cdot P )(z \longi \cdot P) 
-(z \longi \cdot P)^ 2   {P \st ^ 2 \over P \longi ^ 2}            $$
Using (\ref{2.6}) we notice cancellation of the terms proportional to 
 $     (z \st \cdot P )  (z \longi \cdot P)$ and we  can write 
  \beq \widehat Z =  
 z \st ^ 2 P ^ 2 - (z \st \cdot P)^ 2  + 
 P^ 2 (  z \longi ^ 2   -   {  (z \longi \cdot P \longi ) ^ 2 \over P
 \longi ^ 2}  )
     \label{2.7} \eeq
It is convenient to define  the projector  "orthogonal" to $P \longi$, say
\beq \Omega ^ \alpha _ \beta =   \delta ^ \alpha _\beta  - { P \longi ^ \alpha
 { P \longi }_\beta   \over P\longi ^ 2 }     \label{2.8}  \eeq
   because we can write
\beq  z \longi ^ 2    -  { (z \longi \cdot P \longi ) ^ 2  
 \over P \longi ^ 2}   =  (\Omega z \longi )^ 2              \label{2.9} \eeq
and we easily check that  $ (\Omega  z) ^\alpha $  commutes with    
    $ (y \longi \cdot P\longi  )$.
So we  have
\beq \widehat Z = z \st ^ 2  P^ 2 - (z\st \cdot P)^ 2 
   +  (\Omega z \longi )^ 2   P^ 2                     \label{2.10}    \eeq
which justifies the claim that   $ \widehat Z$  
commutes with   $y\longi \cdot P\longi$
Here we notice that 
     $ \Omega z \st  = z \st $ and  
\beq   (\Omega z)^2  = z \st ^2  + (\Omega z \longi)^2    \label {2.10bis} 
                                                                         \eeq
thus  we  finally obtain
\beq \widehat Z   = (\Omega z)^ 2 P^ 2 - (z \st \cdot P )^ 2  
  \label{2.11} \eeq
which is much more tractable than formula (\ref {2.4}).

\medskip
\noi 
Mass-shell constraints can be replaced by their sum and difference, so we set
$$   \mu = \half (m_1 ^2 + m_2 ^2) ,  \qquad \quad
 \nu = \half (m_1 ^2 - m_2 ^2)                      $$
  The  explicit form of   $K'_1$ and  $  K'_2$ 
       was given in Ref. \cite{DVNCim}.                             
  Equations  (3.36) of Ref. \cite{DVNCim}  yield in the present notation  
\cite{novecL}
  \beq K'_1 + K'_2 =
K_1 + K_2  -2T  \   {y \longi \cdot P \longi  \over  P \longi  ^2 }
  + {T^2  \over  ( P \longi ) ^ 2}     \label{2.12}  \eeq      
where 
 \beq T = K_1 -K_2 -  y\longi \cdot P\longi              \label{2.13} \eeq
and the  difference is
\beq  K'_1 - K'_2 = y\longi \cdot P \longi               \label{difK} \eeq
It is noteworthy that  $M_{03} $ and $M_{12}$ are not affected by going over to 
the external-field representation. In other words we can write
\beq     M'_{12} =  M_{12} ,  \qquad    M' _{03} =   M _ {03}  
                                                \label {rotboost}     \eeq
Indeed the transformation from $\Psi$ to $\Psi'$ is formally generated by
$B= LT$ where $L$ and $T$ are given by  (\ref{2.5})     and     (\ref{2.13}) 
respectively \cite{rigo}.
  Commutation of $L$ with $ M_{12}$ and   $  M _ {03}$ is 
obvious.  For commutation of $T$, the only point to be checked is that
 $K_1 - K_2$    actually    commutes with $M_{12}$. But  $K_a = K (a) $ 
where  $K(a)$ is the (half-squared) squared-mass operator  for    particle $a$ 
alone in the field. We know the constants of the motion in the one-body 
sector \cite {bacry}.   In particular we know that $K_a$ commutes with 
both  $ (M_a)_{03}$    and      $ (M_a)_{12} $ .   Thus $T$
commutes with     $  M_ {03}$    and      $ M_{12}$.
Finally $B$ shares the same property, which formally proves (\ref{rotboost}).
Let us prove the following 

\noi {\sl Proposition

\noi   Angular momentum in    $(E\st)$   and   boost     in $(E \longi)$ 
 are constants of the motion\/}.

\noi   In other words we claim that our  squared-mass operators  both
commute with  the transverse and longitudinal components of the total 
angular momentum.

\noi   Working in the external-field representation, all we need is
to prove  that      $  M_ {03}$    and      $ M_{12}$
commute with    both $ K'_a + V'$, or equivalently with 
$K'_1 + K' _2 + 2 V'$ and with   $y \longi \cdot P \longi $. 
Commutation with  $K'_1$ and $K'_2$ separately is ensured from
  the properties of single-particle motion in the field.
Moreover     $y \longi \cdot P \longi $ is invariant under any spacetime 
rotation. The last point to check is whether    $  M_ {03}$    and    $ M_{12}$
actually commute with $V'$. It is sufficient that they commute with all 
arguments of $f$ in formula  (\ref{2.3}), which is true because  these three
arguments  are manifestly Lorentz invariant.

\noi For completeness, it is in order to remind here that pseudomomentum, 
originally represented by 
$$   C= P +  {e\over 2} \  z \cdot F $$
keeps the same expression in the external-field representation ($C' = C$),
and  is also conserved \cite{DVNCim}.

\subsection {Ultimate Representation}
For neutral systems, a further transformation  inspired by the work of Grotch 
and Hegstrom \cite{gro}, and similar to a gauge transformation,
                               permits to get rid of the $Q$ variables.
Transforming the  wave function  yields
$ \Psi '' = ( \exp i\Gamma) \Psi ' $ 
with the help of  the unitary transformation generated by 
  \beq   \Gamma  =  {e \over 2} (z.F.Q)         \label{2.14}        \eeq
We  set 
  \beq {\cal O} ^\sharp  = \exp (i \Gamma) \ {\cal O}  \   \exp (-i \Gamma) 
 \    \qquad      {\cal O} '' =
 ({\cal O}' ) ^ \sharp    \qquad  \forall  {\cal O}         \label{2.15}   \eeq

\noi       The new equations of motion
\beq (H''_1 + H''_2) \   \Psi ''  =  \mu \   \Psi ''    \label{2.16} \eeq
\beq   y \longi \cdot  P \longi    \Psi '' =
  \nu \   \Psi  ''   \label{2.17}   \eeq
may    "look like"  translation invariant, although  {\it they are not}.
The reason is that  pseudomomentum is transformed to $P ^ \alpha $  
by (\ref{2.15}),
that is  $C'' =  P$. Of course $P$ is not any more the generator of spacetime 
translations.  These transformations now have a generator $P''$ which differs 
from $P$ because $\Gamma$ in  (\ref{2.14}) is not translation invariant.
In the ultimate representation considered  here $C''$  generates the 
relativistic analog of the so-called "twisted translations" invoked in 
\cite{avr}.

\bigskip
\noi   From now on 
 {\it we demand that  pseudomomentum be diagonal with a 
timelike four-vector   $k^ \alpha $ as eigenvalue}. Instead of 
 $C^\alp  \  \Psi =  k^\alp \  \Psi$  we are using our ultimate  
representation   and    write   
$  P^\alp   \Psi ''= k ^\alp  \Psi ''$.
Combining this requirement with (\ref{2.17})   we obtain    
\beq  \Psi'' = \exp (i k \cdot Q) 
\ \exp (i \nu { z\longi \cdot k \longi \over  | k \longi | } )\  
 \phi             \label{2.18}   \eeq  
where  $ \phi  $ depends   on $z$,  but only through its 
projection orthogonal to $k \longi $, and additionally depends on $k$ and on 
$\nu$ as parameters. 
In other words $ \phi = \phi ( \nu ,k,  \varpi  z) $ 
with     the following notation.

\noi  {\sl Notation}

For all four-vector $\xi$, we define    
$\varpi \xi  $ as  the  projection  of  $\xi$ onto  the 3-plane orthogonal to 
$k \longi $,  say  
$ \disp (\varpi \xi)  ^\alp = 
\xi ^\alp -  ( \xi \cdot k \longi) \     k \longi ^\alp   /  k \longi ^2     $.

\noi
Similarly   $\xi  \per$ denotes the  projection of  $\xi$ onto the 3-plane
 orthogonal to  $k$. In general  $\varpi z    \not= z \per$, but they 
coincide when  $k \st $ vanishes.

\medskip
\noi It is convenient  to  introduce here the {\it motional parameter\/}
 $ \disp   \epsilon =  {|k\st | \over |k \longi| }$.  When $\epsilon$ doesnot 
vanish, a number of terms involving the contraction  $k \cdot F$ arise. In 
fact    $ (k \cdot F)^\alp =  |k|  \  E ^\alp $ where   $  E ^\alp $  is the 
electric field   "seen" by an inertial observer moving with constant momentum
  $k ^\alp$  ({\it   motional electric field}).
We have  the identity
  \beq  \disp   {1 \over k \longi ^2} = {1 \over k^2} (1 - \epsilon ^2) 
                                                       \label{2.19}     \eeq
Notice that  $k \st$ is linear in $\epsilon$ because we can write
 $k\st =  \epsilon \Lambda  k\longi $ where the second rank 
tensor $\Lambda$ represent the boost from the direction of $k\longi$ to the 
direction of $k\st$ (thus $\Lambda \cdot \Lambda = \delta$).

\subsection{Explicit Formulas}
  The reduced (or internal) wave function $\phi$ must  be determined
 through the "sum  equation" 
 $ (H''_1 + H''_2 ) \   \Psi '' =   \mu   \Psi ''  $, 
 simplified with help of   (\ref{2.18}).      

\noi            Given the function $f$ involved in (\ref{2.2}),
 let us display $H''_1 + H ''_2 $ in detail.
It is clear that 
\beq   H''_1 + H''_2 = 
K''_1 + K'' _2 + 2 V''                               \label{3.1}            \eeq
so we have to transform    $  (K'_1 + K'_2)$  and  $V'$  
according to (\ref{2.15}). We find   that $Q$ and $z$ are unchanged whereas
\beq P ^ \sharp = P  + {e \over 2}  F \cdot z  
\qquad    \quad          P ^\sharp  \longi   =  P \longi    
    \label{3.2}  \eeq   
  \beq y^ \sharp = y  - {e \over 2}  F  \cdot  Q     \label{3.3} \eeq
\beq P^ {\sharp 2} = P^2 + e P \cdot F \cdot z
+ {e^2 \over 4} (F \cdot z)^2                      \label{3.4}  \eeq
\beq (K_1 + K_2) ^\sharp =  { P^2 \over 4} + y^2 
- {e \over 2} z \cdot   F  \cdot P  + {e^2 \over 4}   (z \cdot F) ^2 
                                            \label{3.5}  \eeq
\beq T ^ \sharp =
y \st \cdot P \st - 2e z \cdot F \cdot y    \label{3.6}     \eeq
               \noi
Now we apply  transformation (\ref{2.15}) to (\ref{2.12}),
taking (\ref{3.5})(\ref{3.6}) into account.  It gives 
 $$  K''_1 +  K'' _2 =      $$
\beq          {P^ 2 \over 4} + y^ 2   -  {e\over 2}  z \cdot F  \cdot P 
 + {e^ 2 \over 4}  (z \cdot F )^ 2
+{T^\sharp \over P\longi ^2} (T^\sharp - 2 y\longi  \cdot P \longi)
                                                          \label{3.7}      \eeq
with $T^\sharp $ given by (\ref{3.6}).  
We know that $2 V''$ must be added to this expression in order to obtain
 $H''_1 + H''_2  $.
  But in (\ref{2.14})  $F ^ {\mu \nu} $ is purely transverse, therefore 
   $ (y \longi \cdot P \longi)^ \sharp  = y \cdot P\longi    $.
We have by  (\ref{2.3})
\beq V'' = f( {\widehat Z} ^ \sharp, 
 {P^\sharp }^2,    y \longi \cdot P\longi )   \label{3.8}  \eeq
where $ {P^\sharp }^2 $ is as in (\ref{3.4}) and we must compute
 $\widehat Z ^ \sharp $ from (\ref{2.10}) with   help of (\ref{3.2}).
(We make the convention that 
  $\widehat Z ^ \sharp   = ( \widehat Z) ^ \sharp $ and not the reverse).

\medskip
\noi  To this end we apply the transformation (\ref{2.15}) to eq. (\ref{2.10}).
 A glance at  (\ref{2.9}) 
shows that  $ (\Omega z\longi )^ 2 $ is not affected by the transformation.
Remind that $z$ is unchanged; 
we notice that  $ z \st \cdot P ^ \sharp =  z \st \cdot P $ because,
 $F$ being purely  transverse, 
$ z \st \cdot  F  \cdot z $ identically vanishes.
Thus, using  (\ref{2.10bis}) we obtain
\beq \widehat Z ^ \sharp =
{P^ \sharp}^ 2  (\Omega z )^ 2- (z \st  \cdot  P )^ 2     \label{3.9}  \eeq
Now, eqs (\ref{3.1})(\ref{3.7})(\ref{3.8}) supplemented with
 (\ref{3.4})and (\ref{3.9}) furnish the complete expression of
  $ H''_1 + H''_2 $,  to be inserted into (\ref{2.16}).
At this stage we are in a position to carry out  the reduction.

\bigskip 

\noi   \section{Three-Dimensional Reduction}

\subsection{Calculations}
After transformation to the ultimate  representation we have obtained 
  $C'' = P$.

\noi Calculations can be organized as follows:  
  Whereas (\ref{2.17}) 
fixes the dependence in the relative time, eq.(\ref{2.18}) allows us 
to factorize out the "center-of-mass motion", and we are left with the reduced 
wave function    $ \phi$     which arises in eq. (\ref{2.18}).
Obviously  (\ref{2.17}) implies that  
\beq y \longi \cdot k \longi  \  \phi  = \nu  \      \phi    
     \label{4.1}  \eeq
thus  $\phi$  depends on $z $ only through its projection $ \varpi z$.
 It is clear that $\phi$ generally depends  on $ \nu $ and $ k$ as parmeters. 

\noi
In search for a   reduced wave equation, we  replace $P^ \alpha$ 
 and  $  y \longi  \cdot P \longi $     respectively 
 by  their  eigenvalues  $k^ \alpha $     and  $ \nu $ 
in $H''_1 + H''_2 $,
 and we divide  by exponential factors. 

\noi 
For any operator ${\cal O} $ it is convenient to use the following convention
\beq    ({ \cal O})_{\nu , k } =
{\cal O}   |   _{ y\longi \cdot P \longi = \nu , \    P=k }       
\label{4.2}  \eeq
The subscript $k$ refers to the vector $k$, which finally contributes by its 
longitudinal piece only.
 In this procedure, a term like $y^2$  must be  written  as
    $ \displaystyle  y^2 \equiv  (\Omega y  )^2 +
   { ( y \longi  \cdot  P \longi )^2  \over P \longi ^2 }  $.
If we now introduce the projector $\varpi $ orthogonal to $k \longi $ 
and  use   identity  (\ref{2.19}) we obtain  for instance,  
with help of (\ref{4.1})
\beq   ({P^2 \over 4 } + y ^2 ) _{\nu, k} =
  { k  ^2 \over 4} + (\varpi y)^2  + {\nu ^2  \over k \longi ^2 }     
=    { k  ^2 \over 4} + (\varpi y)^2  + {\nu ^2  \over k^2} 
 - \epsilon ^2 {\nu ^2 \over k ^2 }                \label{4.3}   \eeq 
 which is to be taken into account when computing  
 $ (K''_1 + K''_2) _ {\nu , k } $ from (\ref{3.7}).

\noi                 According to  (\ref{3.1})   we have
$ \displaystyle    ( H''_1 + H ''_2)  _{ \nu , k  } = 
(K''_1 + K'' _2 )_{\nu , k }  +
     2  (V'') _{\nu , k   } $.  
Defining
 \beq  R (\nu, k \longi, k \st) =   (K''_1 + K''_2)_{\nu, k  }  
 \label{4.4}    \eeq
\beq W (\nu, k \longi, k \st) =     (V'')_{\nu,  k }      \label{4.5}   \eeq
Recalling  (\ref{3.1}),   equation  (\ref{2.16}) gets  reduced to
\beq     R    \    \phi  +    2W   \  \phi  =  \mu \phi    \label{4.6}  \eeq
\noi Let us stress  that $\mu $ is just a parameter  fixed from the outset.
As other parameters arise in (\ref{4.6}), namely $k$ and $\epsilon$,
the question wether (\ref{4.6}) can be considered as a spectral problem,
 {\it and for which eigenvalue\/}, is not yet settled and will be considered 
later on, with help of equations (\ref{4.10})(\ref{4.15}). 
See eq. (\ref{4.14}) below.

\medskip
\noi  Since $\phi$ depends on  $z$ only through $\varpi z$, it is important to 
realize that     neither   $R$ nor  $W$  involve  the operator
 $z\longi  \cdot k \longi $. This will be  checked below and will permit us to 
consider equation (\ref{4.6}) as a three-dimensional problem involving 
operators $R$ and $W$ acting on functions of $\varpi z$.

\medskip 
 \noi    The explicit expression of $R$ comes from (\ref{3.7}), with help of 
(\ref{4.4}).  Since $K''_1$ and $K''_2$ are no more than quadratic in the field 
strenght,  let us make the convention that the superscripts $ (1), (2)$ 
respectively refer to  the (homogeneous) linear and quadratic terms in the 
field.  We  start from (\ref{3.7}), compute $K''_1 + K''_2$ to be 
inserted into (\ref{3.1}) and further simplify  with help of 
convention (\ref{4.2}).
The zeroth order contribution to  $R$ is
  \beq R\zer = {k^2 \over 4} + {\nu ^2 \over k\longi ^2}   + (\varpi y)^2   +
y\st \cdot k\st \   {y\st \cdot k\st - 2 \nu \over  k\longi ^2}  
  \label{4.7} \eeq
Applying again identity (\ref{2.19}) 
and  setting 
 \beq (S)_{\nu , k}  = (\varpi y )^2  + 
(y \st \cdot k \st) \  
  { y \st \cdot k \st - 2  \nu    \over  k \longi ^2 }   
- \epsilon ^2  {\nu ^2  \over k^2 }   \label{4.8} \eeq 
we can write
 \beq R   \zer  =  
  {k^2 \over 4 } +  {  \nu ^2 \over k ^2} + (S)_ {\nu, k}       \label{4.9} \eeq
It is convenient to define 
 \beq  \lam = {k^2 \over 4}  + {\nu ^2 \over k ^2} - \mu      \label{4.10}  \eeq
so we can write 
\beq R \zer =     \lam + \mu   + (S)_{\nu , k }        
          \label{4.11} \eeq
The field-depending terms in (\ref{3.7}) provide
   \beq R^ {(1)}=  4 e (z\cdot F \cdot y)    {\nu  \over  k \longi ^ 2 } 
         -  {e \over 2} z \cdot F \cdot k      \label{4.12}     \eeq 
\beq R  ^ {(2)} =
 {e^ 2 \over 4} (z \cdot F)^ 2 
+  4 e ^ 2   { (z\cdot  F  \cdot  y)^ 2  \over k \longi ^ 2 }      
\label{4.13}       \eeq
We remember that $F$ is purely transverse. Contractions involving $F$
only depend on the transverse components;  for instance  $F \cdot k $  is 
just a combination of the quantities  $k\st ^\alpha$.
It is noteworthy that only the transverse components of $z, y $ arise in 
$ \   R  ^{(1)}, \   R ^{(2)}  \   $, whereas $ (S)_ {\nu, k } $  depends on 
  $ \varpi y $ and   $ y \st $.
As a whole,  $R$ depend only on  $ \varpi z $ and  $ \varpi y  $ (recall $y 
\st , z \st $ are pieces of $\varpi y , \varpi z $ respectively).

\noi
 In view of (\ref{4.11})(\ref{4.12})(\ref{4.13}), equation (\ref{4.6}) may 
be finally written
\beq  \lambda \phi + [ (S)_{\nu, k} + R ^{(1)} + R ^{(2)} + 2 W  ] \phi  = 0 
 \label{4.14} \eeq 
The square bracket in (\ref{4.14}) is nothing but $ (-N'')_{\nu k}$ provided,
in the original representation,  we introduce the conserved quantity 
   \beq    -N   = 
{1 \over 4 } C ^2  +    (C^2)^{-1}   (H_1 - H_2)^2   
                     -  (H_1 + H_2)       \label{4.15}              \eeq
now represented by   the operator
  $$-N'' =
{P ^2  \over 4} +  {(H''_1 - H''_2) ^2   \over P ^2} 
  -  (H''_1 + H''_2)                                           $$
and intimately related with the energy of relative motion.

\medskip
\noi  
The last term to be evaluated in (\ref{4.14}) is  $W$.
 In view of (\ref{4.5}) we have first to write down the 
expression for $V''$, say (\ref{3.8}). 
It follows  that 
\beq W= f(   (\widehat Z ^ \sharp )_{\nu, k} , \ 
     (P^ \sharp )^ 2 _ {\nu , k} ,  \   \nu )      
         \label{4.16}    \eeq 
 In this  formula $ (P^ \sharp )^ 2 $ is given 
by (\ref{3.4}) and  $\widehat Z ^ \sharp $ by (\ref{3.9}). 
 Making the substitutions
$P \rightarrow  k$ and  $ y \longi \cdot P \longi  \rightarrow  \nu $,
 hence   $\Omega  \rightarrow  \varpi     $,  we obtain
\beq  (\widehat Z ^ \sharp )_{\nu, k} =
  (P^ \sharp )^ 2 _ {\nu , k} \   ( \varpi z)^ 2 - (z \st \cdot k)^ 2
      \label{4.17}    \eeq
\beq  ( {P^ \sharp}  ^ 2) _ {\nu, k} =   k^ 2 +
e  \    k \cdot F \cdot z    +{e^ 2 \over 4}(F \cdot z)^ 2     
  \label{4.18}  \eeq
It is clear that $W$ does not involve the operator $z \cdot k \longi$.
Formulas   (\ref{4.17}) (\ref{4.18}) are to be  inserted into  (\ref{4.16}), 
then the explicit form of $W$ will come out.

\medskip  
It is natural to consider (\ref{4.14}) as an equation for the eigenvalue 
 $\lam$. 
But we meet a complication because  $\lam$ is not independent from $k^2$.
In fact we can solve (\ref{4.10}) for $k^2$ and insert the result 
      \cite{oneroot}                   into $(N'')_{\nu , k}$. 
As a result (\ref{4.14}) bears a  nonlinear dependence on $\lam$. A similar 
situation was  pointed out by Rizov, Sazdjian and Todorov \cite{rst} in the
 case of isolated systems undergoing energy-dependent interactions.
 In the presence of  magnetic field
however, the reduced wave equation is nonlinear in $\lam$, even in the simple 
case where the mutual interaction term $V\zer$ does not depend on $P^2$.
This can be seen as follows:
first we notice that the occurence of  $ (\widehat Z ^ \sharp )_{\nu, k} $ in 
$W$ brings out a dependence on $k^2, k\longi ^2$. Second we observe an 
unescapable dependence on    $k^2, k\longi ^2$  in formulas 
(\ref{4.8})(\ref{4.12})(\ref{4.13}).
We end  up  with a  nonconventional spectral problem which  requires a 
special  treatment, reserved for a future work.

  \medskip
  \subsection{Discussion}

\noi          Finally the mass-shell constraints have  been reduced to the  
three-dimensional problem of solving (\ref{4.14}). This formula is nonlinear 
in the field strenght and might  be applied to strong fields \cite  {pilk}. 
 Let us review  the various contributions it contains.
We distinguish motional terms, depending on $\epsilon$ or depending on 
$k\st$, where we know that $k\st$ is linear in $\epsilon$.

\noi  Loosely speaking we could say that,
in as much as the shape of  $W$ departs from the original form assumed 
by  $V  \zer$,  every thing goes as if the mutual interaction were somehow
 "modified by the presence of  magnetic field".

\medskip
\noi
{\bf  a) system at rest}

\noi     The  particular case where  pseudomomentum is purely longitudinal
(say $k\st =0$) enjoys a particular simplicity. 
If we assume for a moment that $k$  coincides with $k  \longi$,
it is possible to find a frame where ${\vec k}$ vanishes whereas the 
electromagnetic field is purely magnetic. 
   We refer to this situation as the case  {\it at rest}. 

\noi
In this case, $\varpi z = z \per , \  \varpi y = y \per$ 
and $(S)_{\nu , k} $  simply reduces to $y\per ^2$,
since  $ k\longi $   coincides with $k$.

\noi As $z \st \cdot k$ in  (\ref{4.17})  vanishes, we notice that
                $       (\widehat Z ^ \sharp)   /    
  {P^ \sharp} ^ 2 ) _{\nu, k}  $ 
reduces to  $z \per ^2 $.     According to (\ref{4.16})
 and to a notation defined in Section 2, we can write
$$ W =  g (z \per ^2,  ({P^ \sharp} ^ 2 ) _{\nu, k} ,  \nu)  $$   
where $k \cdot F  \cdot z $ vanishes in (\ref{4.18}),  so 
$  ({P^ \sharp} ^ 2 ) _{\nu, k}  $ reduces to 
$\disp   k ^2  +  {e^2 \over 4} (F \cdot z)^2 $.

\noi  If the mutual interaction doesnot 
depend on the energy, we end up with
$ W =  g ( z \per ^2, \nu) $.
Thus, for energy-independent interactions, namely
 $V\zer = g(Z/P^2, y \cdot P)$,
$W$ assumes the form $g (z \per ^2 , \nu) $. 
 In other words: 

\noi 
 {\em   At rest, the magnetic field doesnot modify the mutual interaction, 
provided this interaction is  not energy-dependent\/}.

\noi
In contrast, if $\   \disp {\partial V \zer  \over \partial P^2} \   $ doesnot 
vanish, the shape of $W$ may  substantially depart from that of  $V \zer$
 in strong fields,  owing to the contribution of 
$(F \cdot z )^2    $ in    $  ({P^ \sharp} ^ 2 ) _{\nu, k}$.
This correction to $V \zer$ is a genuine "three-body" term in this sense that
 it vanishes if either the mutual interaction or the magnetic field is turned 
off (pretending  that the field is generated by a  ficticious 
"third body" located at infinity).

\noi   Looking again at  equation    (\ref{4.14}), we see that,  at rest, 
 all surviving terms not included in $W$ 
can easily be identified as covariant generalizations of the usual terms 
present in the non-relativistic theory  \cite{avr} \cite{herold},
 except for a piece of $R^{(2)}$
which depends on the relative angular momentum,  see contribution of
  $ \disp  { z \cdot F \cdot y  \over  k \longi ^2 }  $
 in formula (\ref{4.13}).
 This  contribution  remains  
small for heavy systems  ($k ^2 >>  F$)  but  might be significant 
for light  systems  ($k^2 <<  F$) in a strong magnetic field.

\medskip
\noi   At {\it first order in the field strenght\/} however,
 the  relative motion admits no 
correction other than a term proportional to $\nu$ (indeed $F \cdot k $ 
vanishes). For equal masses this term is zero and  there is no departure
 from the motion of an isolated system.

\medskip
\noi    {\bf   b)  motional case}

\noi
When $k\st$ is nonzero, we reckognize the  motional electric field contained 
in $z   \cdot F \cdot k$.                                           
 For energy-dependent potentials, and even in a  weak field, this term 
contributes to  $W$  through  (\ref{4.16}). 
  But of course, it may be  neglected in case of slow motion in a weak field,
 where both $\epsilon$ and $F$ are considered as first order quantities, 
which  entails that  $F \cdot k$  is  a second order quantity.
On the one hand, this can be seen as a stability property of the   
neutral two-body system, under application of a constant field.
 But on the other hand, it forces one to go beyond the 
weak-field-slow-motion approximation if one wishes to compute significant 
corrections to the energy associated with  relative  motion.

\bigskip
            \section{Conclusion}
                           
\noi      The coupled Klein-Gordon equations describing a globally neutral 
system  have been  reduced to a three-dimensional  equation
involving truly motional terms and recoil effects in a   covariant fashion.
In this formulation  the particular symmetry associated with a constant
 magnetic field {\it in space-time\/}  is manifestly respected.
After separation of the internal motion, and after factorizing the 
dependence on relative time, the surviving number of degrees of freedom is 
finally the same as in the nonrelativistic theory.

\medskip 
\noi
 We  now have  a clean theoretical basis for the   study  of relativistic 
 bound states in a constant magnetic field, the  simplest of all the  cases
 where  an external field is present. 

\noi
In the reduction procedure it was essential to consider eigenstates of 
pseudomomentum.
The square of this vector plays the role of an effective squared mass 
which can be, in principle, evaluated by solving the reduced wave equation.
But the eigenvalue problem involved in this equation  is crucially
  non-conventional, for the eigenvalue arises in a nonlinear way, 
 even if  mutual interaction doesnot depend on the total energy.
This situation  requires a refinement of conventional methods; the method  
devised  in Ref. \cite{rst} will  help to carry out this task in the future. 

    \noi
Our formulas   are quadratic  in the field strenght and offer
 a starting point for investigating  strong field effects.
In principle, they encompass all kinematic possibilities of the system as a 
whole and   permit a description of ultra-relativistic situations, where 
$|k \st|^2    \simeq  |k \longi |^2   $.

\noi
In the present state of the art, we notice that, in a {\em weak} field,
 the {\em slow} collective motion (first order in $\epsilon$) of opposite 
charges interacting through 
a potential which doesnot depend on the energy, escapes the above-mentioned 
complication; but in this case the presence of external field results in a 
first order Starck effect which obviously vanishes for generic shapes of the 
mutual interaction potential.  
For the harmonic oscillator for instance, this remark indicates that 
 the naive quark model enjoys some kind of stability property.
But if we have perturbation theory in mind,
the computation of significant corrections requires the  setting of a 
 nonconventional treatment.

\medskip
\noi
In sofar as approximations are concerned,  it is in order to realize that 
two situations are possible:

\noi
Either the magnetic field is considered (like in the previous example) as a 
perturbation applied to the system. 
Or, in contrast,  the mutual interaction is treated as a perturbation like in 
 the helium atom.

\noi
In that latter case, the zeroth order approximation describes two independent 
particles  moving in the magnetic field; in this unperturbed motion, the 
transverse degrees of freedom are bound by the magnetic field
(corrections to the corresponding spectra are reserved for future work).
We expect to avoid the pathology of "continuous dissolution"
 \cite{dissol}  \cite{brok}  for two reasons:
 The particules we consider here have no spin,
and we  can impose positive individual energies,
requiring that both $P \cdot p_1$ and $P \cdot p_2$ have  positive eigenvalues.

\medskip
\noi The Ansatz which allows for a three-dimensional reduction in our covariant 
framework  automatically generates various terms in the wave equation.
We have seen  that the importance of these terms depends on the strenght of 
the field and on the  state of motion of the system as  a whole.
Inspection of these terms  indicates that, from a practical point of view,
 the shape of the mutual interaction is "somehow modified" 
by the magnetic field.
 As can be read off from  (\ref{4.18}),  the modification implied by 
 (\ref{4.16}) is quadratic  in $F$  and may become dominant in  strong fields
provided $V\zer $ is energy-dependent. 
This point  concerns most of the realistic two-body potentials.              

\noi     Further work is needed  in order to get beyond these qualitative
 indications. 

    \noi
For the sake of simplicity we have focused here on scalar particles,
 but naturally  an extension to particles with  spin is desirable.
A generalization to globally charged systems would also be of interest.
    
\noi
Let us finally mention that, in principle, the contact with more conventional
methods of quantum field theory could be improved, trying to directly derive
 all our terms from  a  Bethe-Salpeter  equation {\em that  takes  the magnetic
 field into account  from the start\/}.       This would 
    mean to remake the  work of Bijtebier and Broeckaert \cite{brok} in a
 way which respects the particular symmetry of constant magnetic field, $ i.e.$ 
    treating all the possible lab frames on the same footing.                
To our knowledge, nobody has yet carried out this task.


\begin{thebibliography}{3}
\bibitem {purmagn} We say that a field is pure electric (resp. pure magnetic) 
when there exist frames where the field appears to be so; of course these two 
situations exclude one another.
\bibitem {novecL} To prevent confusion, we now avoid using the vector
 $L^\alp$ introduced in \cite {DVNCim}, because $L \cdot L$ was distinct from 
the square of the scalar function $L$.
\end{thebibliography}
\end{document}